\documentclass[reqno]{amsart}
\usepackage{graphicx}
\usepackage{citesort}

\hoffset = - 1.75 cm
\begin{document}

\title{On energy for accelerating observers in black hole spacetimes}

\author{Seth A. Major}

\date{April 2022}
\address{Department of Physics\\
Hamilton College\\
Clinton NY 13323 USA}

\email{smajor@hamilton.edu}

\begin{abstract}
Energies for constantly accelerating observers in Ba\~nados, Teitelboim, and Zanelli (BTZ), Schwarzschild and Schwarzschild-de Sitter spacetimes are derived. The expressions are in terms of acceleration, cosmological constant, and area, quantities measurable by the observers.  Based on results from quantum fields in curved spacetime for the redshifted Hawking temperature, quasi-local entropy and thermodynamic-like laws are briefly explored in the three spacetimes.
\end{abstract}

\maketitle

The definition of energy is notoriously subtle in gravitational physics. Although no notion of local energy exists in general spacetimes, there are useful notions of quasi-local energy for regions of spacetime (for a reviews see \cite{szabados,MassRev}.) Choices such as slicing and boundary conditions have led to a plethora of possible definitions quasi-local energies and masses, e.g. \cite{penroseQE,hawking,geroch,KO,JK,DM,B,NTJ,R,hayward,BY,BLY,lau,RT,HH,HM,MSM,HaywardEnergy}.
These choices are aspects of formulation: Defining energy is a matter of choice and, in particular, physical characterization: {\em of what} and {\em for whom} are we defining energy. Considerations in that characterization are as often practical as foundational: Capturing the energy content in gravitational waves suggests different choices than definitions of energy best suited to static or stationary spacetimes.  More broadly, in statistical models the energy of frozen degrees of freedom is rightly neglected. Choosing the `correct' energy is often a matter of selecting the best approximate form to capture the relevant physics. Additionally in the context of general relativity,
``...the formulation of meaningful global or quasi-local mass and angular momentum notions in General Relativity {\em always} needs the introduction of some additional structure in the form of symmetries, quasi-symmetries or some other background structure" \cite{MassRev}.
This note defines an energy for constantly accelerating observers (CAOs) in black hole spacetimes with a timelike Killing vector field.\footnote{I assume that these observers are not unlike us.  In addition to the usual garden variety meter sticks and clocks, I assume that they have access more sophisticated equipment e.g. accelerometers, radar or laser ranging, bolometers, and telescopes.}  
This energy is `quasi-local' in the sense that there is a sufficient number of observers positioned on a surface to perform the measurements rather than, for instance, a curvature expression integrated over the bounding surface of a compact spatial region. This choice of observers limits the generality of the resulting structure but the hope is that the operational perspective will yield a new tools and views into statistical models of black hole spacetimes. 

The spacetimes considered here all have a timelike Killing vector field, $\xi^a$. The existence of this symmetry offers enough structure to tie a local definition of mass to the asymptotic (`ADM') mass, when such an asymptotic region exists.  In these spacetimes the Killing field has unit normalization, $\xi^a \xi_a = -1$ in the asymptotically flat region, so that $\xi^a$ coincides with the 4-velocity of inertial observers there. Due to Killing's equation and the geodesic motion of a particle in free fall with momentum $p^a$, the energy ${\mathcal E} = - p_a \xi^a$ is conserved along the particle's history.  A local observer with 4-velocity $u^a$ measures the energy of this particle to be $E= - p_a u^a$ anywhere in the spacetime.  Relying as it does on the relative motion of the observer and particle, this expression may not seem to be a promising expression to learn about the energy of the spacetime \cite{alex}. Nevertheless, this expression can capture relevant aspects of the spacetime when the observers' 4-velocity of these accelerating observers is proportional to the Killing field $\xi^a$, $u^a = \xi^a/V$ with the `redshift factor' $V= \sqrt{-\xi^a \xi_a}$. Given these relations the change in the locally measured energy $dE$ due to the flow of energy-momentum by the observers' surface is proportional to the conserved energy $d{\mathcal E}$, 
\begin{equation}
	\label{Eloc}
	dE= - p_a u^a = - p_a \frac{\xi^a}{V} = \frac{d{ \mathcal E}}{V}.
\end{equation}
This relation ties the observers' local energy $dE$ to both the conserved energy $d {\mathcal E}$ and the spacetime geometry through $V$. For example, when considering black holes accumulating mass/energy $dM$ infalling from far away, we can consider changes to the local energy, $dE = dM/V$.\footnote{Directly integrating this expression gives the Brown-York quasi-local energy.}
Of course, this construction is made possible by the existence of the timelike Killing vector field and depends on the choice of norm of $\xi^a$. This energy is also well-known, often called the ``shell" energy. 
One can arrive at this simple form of energy $dE = dM/V$ by considering the conserved flux $\delta T^a_b \xi^b$ of test matter fields with stress-energy tensor $\delta T^{ab}$, as shown in appendix \ref{Enorm}.

The strategy to define the (quasi-)local energy for these accelerated observers is the following: In each class of spacetime geometry the locally measured energy differential of equation (\ref{Eloc}) is re-expressed in terms of the magnitude of the constant acceleration $g$. Although it is technically easier to integrate the areal radius, conceptually this expression is integrated from zero mass to the final mass $M$. The resulting expressions, denoted $E_g$, are the locally measured energies in the spacetimes. In every case the energy is expressed in terms of measurable quantities accessible to the CAOs. These energies for constantly accelerating observers in the black hole spacetimes are the primary results of this work. 

To explore possible (quasi-)local thermodynamics for these observers, possible expressions for local entropy functions $S_g$ are derived based on the expressions for temperature, assuming a first law.  In the different spacetimes, the local temperatures $T_g$ from the literature are a necessary input in the derivation of the local entropy $S_g$. In a first look into the viability of these expressions, consistency checks are performed such as third law and additivity properties.  These entropies are compared to existing entropy expressions. Additionally, in each spacetime one can check that the mass integration at constant radius yields the Brown-York quasi-local energy, which is given by the spatial integration of the trace of the extrinsic curvature $k$, compared to a reference spacetime curvature $k|_o$,  \cite{BY}
\begin{equation}
	\label{EBY}
	E_{BY} = \frac{1}{8 \pi} \int_{\partial \Sigma} (k- k|_o) \sqrt{\sigma} d^2x, 
\end{equation}
where $\sigma_{ab}$ is the induced metric on the two-surface $\partial \Sigma$. This constant-radius energy is denoted $E_r$.   

Part of the motivation for this work arises from the first law of black hole mechanics,
\[
	\delta M = \frac{\kappa}{8 \pi} \delta A_{\mathcal H} + \Omega  \delta J.
\]
Somewhat curiously, this thermodynamic description employs quantities that span the spacetime from the horizon, with quantities of surface gravity $\kappa$ and area $A_{\mathcal H}$, to quantities well defined in `asymptopia' (the asymptotic region), with quantities of the mass $M$ and angular momentum $J$.  This note initiates work on whether there is a meaningful version of this law using (quasi-)local quantities centered on the quantities acceleration, area, etc., accessible to CAOs.

There are several threads of work closely related to the approach taken here.  For external observers in the near-horizon limit the membrane paradigm of black holes (see \cite{thorneMP} and references therein) offers a view on the rich structure of electromagnetic fields and spacetime geometry for observers perched just outside the black hole. In spherical symmetric spacetimes prior work on definitions of energy or mass include the Misner-Sharp-Hernandez mass \cite{MSM,HaywardEnergy}. In constructing statistical models of near horizon geometry Frodden, Ghosh, and Perez \cite{FGP} find striking simplicity in the form of the near-horizon energy and resulting statistical mechanics. In exploring a similar energy, M\"akel\"a pointed out the role of changing mass for constantly accelerating observers \cite{makelaThermo,makela}. 

A perspective on the construction of energy can be seen from Newtonian mechanics. A constantly accelerating observer outside a mass $M$ has an acceleration with magnitude $g_N = |\vec{g}| = M/r^2$.  (Units are such that $c,G,$ and $k$ are set to 1. ) Since this acceleration is constant, $d g = 0$, $dM = (2M/r) dr$. As a bit of mass $dM$ is added to, or removed from, the central mass $M$ we have
\begin{equation}
	dM = \frac{M}{r^2} 2 r dr = \frac{g_N}{4 \pi} dA,
\end{equation}
with change in area $dA$ at the observers' radius.  Integrating gives $M = g_N A/4 \pi$, which can also be found directly from the expression for $g$.  The result holds everywhere away from the origin. The remainder of this note generalizes similar analyses for $E_g$ in black hole spacetimes.

In the next section energies and entropies in the non-rotating and rotating cases of Ba\~nados, Teitelboim, and Zanelli (BTZ) spacetimes are derived. In section \ref{schwarzschild} the analysis is repeated in Schwarzschild spacetimes. The energy and entropy in Kottler (or Schwarzschild-de Sitter) spacetimes are explored in section \ref{kottler}.  The energy is found but the lack of an asymptotically flat region complicates the derivation of the entropy. Kottler spacetimes have a spacetime characterizing mass/energy parameter (SCM) defined as that mass which characterizes the influence of an isolated object on distant (but not too distant or the gravitational effects of the black hole may be occluded by the background spacetime) test objects. Following the paradigmatic example of the Newtonian limit of the Schwarzschild, I denote this mass as $M$ throughout, although in the cases where asymptopia consists of (anti-)de Sitter spacetime, the grounding in asymptotic Minkowski space is not available.\footnote{I have the case in mind in which an effective asymptotia is ``far away" but not at infinity so that there is an effective Newtonian limit between the black hole region and the surrounding cosmological region. For spacetimes with cosmological constant this means $M < 1/\sqrt{\Lambda}$.} In each of these spacetimes an energy for CAOs is defined and the energies are all expressed in terms of quantities that CAOs may measure.   The simplest derivation in general relativity is that of a black hole in $(2+1)$-dimensions.

\section{BTZ spacetimes}

Despite the lack of local degrees of freedom, thirty years ago Ba\~nados, Teitelboim, and Zanelli showed that a black hole solution in $(2+1)$-dimensional gravity exists when there is a cosmological constant \cite{btz}. These spacetimes have constant curvature, neighborhoods of every event are isometric to anti-de Sitter (adS) spacetime, and the solution is asymptotically anti-de Sitter black hole with cosmological constant $\Lambda = - 1/\ell^2$ \cite{btz,carlip95rev}.  In Schwarzschild-like coordinates, the $(2+1)$-dimensional spacetime metric is
\begin{equation}
	\label{btzmetric}
	ds^2 = - N^2 dt^2 + N^{-2} dr^2 + r^2 \left(d\varphi + N^\varphi dt \right)^2
\end{equation}
with
\begin{equation}
	N = \left( -8 M +\frac{r^2}{\ell^2} + \frac{16 J^2}{r^2} \right)^{1/2} \text{  and } N^\varphi = - \frac{4 J}{r^2} \text{  with } |J| \leq M \ell. 
\end{equation}
The cosmological constant provides the required physical scale for a horizon.\footnote{Since Newton's constant, the speed of light, and mass do not create a length in $(2+1)$-dimensional gravity, the model must have an additional scale \cite{SSP}.}  There are black hole solutions that have two horizons, an outer horizon, $r_+$, and an inner Cauchy horizon, $r_-$, where
\begin{equation}
	r_\pm^2 = 4M \ell^2 \left[1\pm \sqrt{1- (J/M\ell)^2} \right].
\end{equation} 
The mass and angular momentum are related to these radii as
\begin{equation}
	M=(r_+^2 +r_-^2)/8 \ell^2 \text{ and } J = r_- r_+/4 \ell.
\end{equation}
Due to the dimensional reduction and lack of local degrees of freedom BTZ black hole spacetimes are considerably simpler than higher dimensional black holes. Local temperature \cite{DL,HL,bcm}, boundary terms \cite{btz,bcm}, and entropy \cite{strominger,bss,carlip_bms} have been determined.  

In the non-rotating case ($J=0=r_-$) CAOs outside the black hole, when $r>r_+$, have an acceleration $a^b = u^c \nabla_c u^b$ with magnitude, $g = \sqrt{a_b a^b}$, 
\begin{equation}
	\label{g_btz_nonrot}
	g = \frac{ r}{\ell \sqrt{r^2 - r_+^2}}.
\end{equation}
As expected, this proper acceleration diverges near the horizon; $g \simeq (\ell \sqrt{2 \epsilon})^{-1}$ when $r=r_+(1+\epsilon)$, $\epsilon \ll 1$.  In the asymptotic limit $r \rightarrow \infty$, the acceleration is approximately $g\simeq 1/\ell$, its minimum value. The Killing vector field $\xi^\mu = \left( \partial_t \right)^\mu$ has norm $\sqrt{-\xi_\nu \xi^{\nu}} = N=V = \ell/\sqrt{r^2-r_+^2}$.

To obtain the energy measured by CAOs for a black hole of mass $M$ one can integrate over mass $m$ from 0 to $M$, keeping the acceleration fixed,
\begin{equation}
	E_g = \int_0^M \frac{\left. dm\right|_g}{V}.
\end{equation}
Since the acceleration of equation (\ref{g_btz_nonrot}) is constant, it is useful to express the SCM $M = M(g,r)$ in terms of the acceleration,
\begin{equation}
	\label{btz_mass}
	M = \left( \frac{r^2}{8 \ell^2} \right) \left( 1 - 1/g^2\ell^2 \right).
\end{equation}
We can see that, at finite $r$, the mass goes to zero as $g \rightarrow 1/\ell$. The mass remains finite in the asymptotic limit if $r \sqrt{ 1 - 1/g^2\ell^2} = r_+$ remains finite.

Constantly accelerating observers have $dg=0$ and so
\begin{equation}
	\left( \frac{\partial M}{\partial r} \right)_g = \frac{2M}{r} = \left( \frac{r}{4 \ell^2} \right) \left( 1 - 1/(g \ell)^2 \right).
\end{equation}
The change in energy measured by these observers is most easily compared to the changing circumferential radius $r$,
\begin{equation}
	\label{dEbtz_norot}
	dE_g = \frac{dm|_g}{V} =  \frac{1}{V} \left( \frac{ \partial m}{ \partial r} \right)_g dr = \left( \frac{g}{2 \pi} \right) \left( \frac{2 \pi dr}{4} \right) \left( 1 - \frac{1}{(g\ell)^2} \right).
\end{equation}
The integration is immediate giving 
\begin{equation}
	\label{Ebtz_norot}
	E_g = \int_0^r d E_g = \left( \frac{g}{2 \pi} \right) \left( \frac{2 \pi r}{4} \right) \left(1 - 1/(g \ell)^2 \right),
\end{equation}
the energy for CAOs at circumferential radius $r$. The expression for the energy $E_g$ may be expressed in terms of the acceleration, local geometry, and the cosmological constant.

The local temperature may be found using quantum fields on curved spacetime techniques. In the case of BTZ black holes, the quotient construction in higher dimensions significantly simplifies the computation of both Green's functions (see e.g. \cite{carlip95rev} and references therein) and the transition rate formula for Unruh-DeWitt detectors \cite{HL}. To compute these quantities one must choose a vacuum state for the field, such as the Hartle-Hawking state which is thermal far from the black hole. In particular, Hodgkinson and Louko consider Unruh-DeWitt detectors and a massless conformally coupled scalar field in a Hartle-Hawking vacuum state with a complete slate of boundary conditions at the asymptotically anti-de Sitter spatial infinity \cite{HL}.  
The Unruh-DeWitt detector approach provides an operational definition of the particle content of the configuration. Generally one finds for this case that the observers measure a redshifted temperature given by the surface gravity $\kappa$, where $\xi^a \nabla_a \xi^b = \kappa \xi^b$ on the horizon, and the redshift factor $V$, sometimes called the Tolman factor.  In the non-rotating case \cite{LO,SM,carlip95rev,bcm}
\begin{equation}
	\label{BTZtemp}
	T_g = \frac{\hbar \kappa}{2 \pi} \frac{1}{V} = \frac{ \hbar r_+}{2 \pi} \frac{1}{\sqrt{r^2-r_+^2}} = \frac{ \hbar g}{2 \pi} \left( 1 - \dfrac{1}{g^2\ell^2} \right)^{1/2}.
\end{equation}
As expected, this temperature diverges with the acceleration near the black hole horizon. The temperature vanishes when the acceleration reaches its minimum value of $1/\ell$, due the vanishing mass in this limit.

A candidate expression for the local entropy may be obtained from a local form of the first law 
\begin{equation}
	\label{1st_btz}
	dE_g = T_g dS_g.
\end{equation}
From equations (\ref{dEbtz_norot}) and (\ref{BTZtemp}) this suggests 
\begin{equation}
	dS_g = \frac{2 \pi dr}{4 \hbar} \left(1 - 1/(g \ell)^2 \right)^{1/2}
\end{equation}
which integrates to
\begin{equation}
	\label{btz_entropy}
	S_g = \left( \frac{2 \pi r}{4 \hbar} \right) \sqrt{ 1 - 1/(g \ell)^2 }.
\end{equation}
This entropy is related to the local circumference as $S_g = (A/4 \hbar) \left(1 - 1/(g \ell)^2 \right)$, where the leading term remains in the  Bekenstein-Hawking ``$A/4$" form, where now the `area' is circumference at the observer's radius, $A=2 \pi r$, rather than at the horizon $r=r_+$.

As a consistency check one may note that as a function of the energy the entropy is 
\begin{equation}
	\label{btzSE}
	S_g = \frac{2 \pi}{\hbar g } \frac{E_g}{\sqrt{1- 1/(g \ell)^{2}}}.
\end{equation}
The temperature from $T = \left( \partial S /\partial E \right)^{-1}$, is equal to $(\hbar g/2 \pi) \sqrt{1 -1/(g \ell)^2}$, as in equation (\ref{BTZtemp}). 

Some additional thermodynamics-like relations hold: The temperature is trivially constant due to the choice of observers. As one can see from $E_g$, as the acceleration reaches its minimum value (and $T\rightarrow 0$), the entropy vanishes. Although whether the entropy is nondecreasing in physical processes remains to be seen, the simple proportionality of equation (\ref{btzSE}) is promising: the entropy satisfies additivity and is a nondecreasing function of $E_g$. 

In this case, the entropy $S_g$ is equivalent to the the Bekenstein-Hawking entropy. (This will not be the case in higher dimensions.) Presumably due to the lack of local degrees of freedom, the entropy (\ref{btz_entropy}) is also a simple re-expression of the Bekenstein-Hawking entropy on the horizon $r_+$, $S_{BH}=A_+/4 \hbar$, where $A_+ = 2 \pi r_+$ is the circumference of the horizon.\footnote{The entropy is ``constant" in the sense that $S_g = S_{BH}$ for all $g$.} To see this, one may use the relation between the circumferential radius and the acceleration, $r = r_+ / \sqrt{ 1 - 1/(g\ell)^2}$. 
 
One may also find a possible expression for the entropy by calculating the circumference in the global embedding Minkowski spacetime (GEMS) construction. In the GEMS construction the entropy is determined by the area found in the higher dimensional embedding spacetime \cite{DL}. (For work on the applicability of the GEMS construction see \cite{paston}.) This is done in appendix \ref{EGEMS} for these accelerating observers with the result that the circumference is
\begin{equation}
	\label{btz_area}
	A_{GEMS} = 2 \pi r \left( 1 - 1/(g\ell)^2 \right)^{-1/2},
\end{equation}
which has a limiting value of $A_+ = 2 \pi r_+$ as $g\rightarrow \infty$. The circumference $A_{GEMS}$ diverges as $g$ reaches its minimum value, which is expected from the geometry as the area in the higher dimensional space becomes unbounded in adS$_3$ \cite{DL}.  However, were we to generalize the Bekenstein-Hawking entropy in the form $A_{GEMS}/4$ then the entropy of these observers would also diverge, contrary to third-law-like behavior. 

As can be seen through the quasi-local constructions of Brown and York \cite{BY}, work terms might arise. In this case however, choosing adS$_3$ as the reference spacetime and identically accelerating observers in adS$_3$, the pressure vanishes. This is briefly discussed in appendix \ref{pressure}. 

We can also integrate the expression for the energy keeping the radius - and not the acceleration - fixed. Integrating the energy $d E_r = dm|_r/V$ with respect to mass $m$ from 0 to $M$ one obtains the energy at constant radius
\begin{equation}
	E_{r} = \int_0^M  \frac{dm}{V} = \int_0^M \left( \frac{\ell}{r} \right) \left( 1- \frac{8 m \ell^2}{r^2} \right)^{-1/2} dm = \left( \frac{2r}{\ell} \right) \left( 1 - \sqrt{1 -\frac{r_+^2}{r^2} } \right),
\end{equation}
which is the Brown-York quasi-local energy $E_{BY}$ of equation (\ref{EBY}) for observers at fixed $r$, when the background is chosen so that $E_{r} = 0$ when $M=0$ (when $\epsilon_o = -1/\pi \ell$ in the notation of \cite{bcm}).

In the rotating ($J\neq0$) case, the literature contains results on the temperature for exterior observers rigidly co-rotating with the horizon, having a (coordinate) angular velocity of $\Omega = r_- / (r_+ \ell)$. (There is evidence both in the Unruh-DeWitt detector \cite{HL} and GEMS \cite{DL} approaches that there is not a well-defined thermal state for stationary observers in the general rotating case.)  With these results for the temperature, the derivation of the energy and possible entropy focuses on the case of rigidly co-rotating observers.

The co-rotating observers' motion satisfies $d \varphi /dt = \Omega$  and they follow the Killing vector field $\chi^\mu = (\partial_t)^\mu + \Omega (\partial_\varphi)^\mu$.  Using angular velocity $\Omega$ in the metric of equation (\ref{btzmetric}), thereby adapting the metric for these observers, makes the calculation of the acceleration, $g = \sqrt{a_b a^b}$, straightforward. The result is 
\begin{equation}
	\label{btz_rot_g}
	g = \frac{1}{\ell} \sqrt{ \frac{r^2-r_-^2}{ r^2 - r_+^2}}.
\end{equation}
As in the non-rotating case, the acceleration diverges at the horizon and has a minimum value of $1/\ell$ far from the black hole. 

The calculation of energy $d E_g = (1/V) d m|_g$ 
follows similar steps as in the non-rotating case. The constant acceleration means
\begin{equation}
	dm|_g = - \frac{\partial g / \partial r}{\partial g / \partial m} dr - \frac{\partial g / \partial J}{\partial g / \partial m} dJ.
\end{equation}
The change in energy becomes
\begin{equation}
	\label{Egbtzrot}
	\begin{split}
	d E_g &= \left( \frac{g}{2 \pi} \right) \left( \frac{ 2 \pi dr }{4} \right) \left( 1 - \frac{1}{(g \ell)^2} \right) \left( \frac{ 1 - \Omega^2/g^2}{ 1 + \Omega^2/g^2} \right) \\
	 &+ \Omega \frac{ \left(1+g^2 \ell^2 \right) \sqrt{(1-1/(g \ell)^2)(1- (\Omega / g)^2) }}{r_+\left(1 - (\Omega \ell)^2 \right) \left(1 + (\Omega/g)^2 \right)} d J.
	 \end{split}
\end{equation} 

When the observers are rigidly co-rotating with the horizon, Hodgkinson and Louko \cite{HL} find that the transition rate of their Unruh-DeWitt detectors is thermal (in the sense of satisfying the Kubo-Martin-Schwinger (KMS) condition) with the local temperature of
\begin{equation}
	\label{tempbtzrot}
	T_g= \frac{\hbar}{2 \pi l} \sqrt{ \frac{r_+^2 - r_-^2}{r^2 - r_+^2}} = \frac{\hbar g}{2 \pi} \sqrt{1 - 1/(gl)^2}, 
\end{equation}
which, in terms of $g$, is the same as the non-rotating case. This expression agrees with the temperature derived with the GEMS construction \cite{DL}.

A first law of the form $dE_g = T_g dS_g - \tilde{\Omega} dJ$, where $\tilde{\Omega}$ is 
\begin{equation}
	\tilde{\Omega} = - \Omega \frac{ \left(1+g^2 \ell^2 \right) \sqrt{(1-1/(g \ell)^2)(1- (\Omega / g)^2) }}{r_+\left(1 - (\Omega \ell)^2 \right) \left(1 + (\Omega/g)^2 \right)}, 
\end{equation}
from equation (\ref{Egbtzrot}). This suggests that the integrated entropy is 
\begin{equation}
	S_g = \left( \frac{ 2\pi r}{4 \hbar} \right) \left( 1 - \frac{1}{(g \ell)^2} \right)^{1/2} \left( \frac{ 1 - \Omega^2/g^2}{ 1 + \Omega^2/g^2} \right).
\end{equation}
This entropy has similar properties to the entropy in the non-rotating case. The apparent vanishing when $g \rightarrow \Omega$ would occur on the inner horizon, when $r=r_-$, which is inside the region under consideration. As shown in appendix \ref{EGEMS}, the GEMS construction gives the same expression for the area as before.

If one ensures that both the acceleration and the angular velocity are constant then $dJ = - J/M \, dM$ and the energy may be integrated yielding
\begin{equation}
	\label{Egbtzrot2}
	E_g = \left( \frac{g}{2 \pi} \right) \left( \frac{ 2 \pi r }{4} \right) \left( 1 - \frac{1}{(g \ell)^2} \right)
	\left[ \frac{ 1 - \left( 1-1/(g \ell)^2 - \Omega^2/g^2 \right) \Omega^2 \ell^2}{ 1 + 3\left( 1 + 1/(g \ell)^2 + \Omega^2/(3 g^2) \right) \Omega^2 \ell^2} \right].
\end{equation}   
The entropy in this case is this expression divided by the temperature in equation (\ref{tempbtzrot}),
\begin{equation}
	S_g = \left( \frac{ 2\pi r}{4 \hbar} \right) \left( 1 - \frac{1}{(g \ell)^2} \right)^{1/2} \left[ \frac{ 1 - \left( 1-1/(g \ell)^2 - \Omega^2/g^2 \right) \Omega^2 \ell^2}{ 1 + 3\left( 1 + 1/(g \ell)^2 + \Omega^2/(3 g^2) \right) \Omega^2 \ell^2} \right].
\end{equation}

\section{Schwarzschild spacetimes} 
\label{schwarzschild}

The Schwarzschild metric is
\[
	ds^2 = - \left( 1 - \frac{2 M}{r}  \right) dt^2 + \left( 1 - \frac{2 M}{r} \right)^{-1} dr^2 +r^2 d \Omega_2^2.
\] 
The norm of the timelike Killing vector field gives $V = \sqrt{-\xi^a \xi_a} = \sqrt{1- 2M/r}$. For constantly accelerating observers as the asymptotic mass of the spacetime changes, the radial location of the observers changes as well. As before, it is useful to consider variable mass $m$ as a function of radius. Since the magnitude of the proper acceleration,
\begin{equation}
	\label{g}
	g = \frac{m}{r^2} \left(1 - \frac{2m}{r} \right)^{-1/2} = g_N \left(1 - \frac{2m}{r} \right)^{-1/2}
\end{equation}
is constant, the mass changes as
\begin{equation}
	\frac{m}{r} =  u \left( \sqrt{1+u^2} -u \right) \text{  with } u =gr.
\end{equation}
(This is the positive and physical root of the quadratic, $m^2 + 2g^2 r^3 m - g^2 r^4 =0$.)  Similarly the measure $dm|_g$ becomes \cite{makela}
\begin{equation}
	dm|_g =  \frac{m}{r} \left( \frac{ 2r - 3 m}{r-m} \right) dr
\end{equation}
and the integration of the energy $dE_g = dm|_g/V$ becomes
\begin{equation}
	E_{g} = \int_0^{r_g(M)} \left( 1 - \frac{2 m}{r} \right)^{-1/2} \left( \frac{m}{r} \right) \left( \frac{ 2r - 3 m}{r-m} \right) dr.
\end{equation}
It is handy to re-express the masses in terms of $u = \sinh x$ and note the following
\begin{equation}
	 \left( \frac{ 2r - 3 m}{r-m} \right) = 2 - 
	 \tanh x. 
\end{equation}
The resulting integration yields
\begin{equation} 
	\label{intE2}
	E_{g} =  g \frac{ A}{4 \pi} - \frac{1}{2g} \left[ \ln \left(\sqrt{1+(gr)^2} - gr \right) + gr \sqrt{1+(gr)^2} \right].
\end{equation}
The quasi-local energy $E_g$ for constantly accelerating observers in Schwarzschild spacetime can be expressed in terms of the areal radius of the CAOs, $r = \sqrt{A/ 4 \pi}$ and the acceleration.  
In the high acceleration, near-horizon limit, the energy simplifies, reaching the expected $E_g \simeq g A/ (8 \pi)$. This limiting energy can be written in terms of the proper distance to the horizon $d$, since $g$ and $1/d$ diverge in the same manner near the horizon; $E_{g} \simeq A/ ( {8 \pi d})$ \cite{FGP}. When the acceleration is low, far from the black hole, the energy goes to $E_g \simeq g r^2 \simeq M$ as expected.\footnote{In the Newtonian limit when $r \gg 2M$ we have $g \simeq g_N \left( 1 + M/r \right)$. Expanding,
\[
	E_g \simeq  \frac{g_N A}{4 \pi} \left( 1 + \frac{g_N \sqrt{A}}{3 \sqrt{\pi}} \right) = M\left( 1+ \frac{2}{3} \frac{M}{r} \right).
\]} 
Other ways to express $E_g$ are discussed in appendix \ref{altEgS}.

\begin{figure}
	\includegraphics[scale=1]{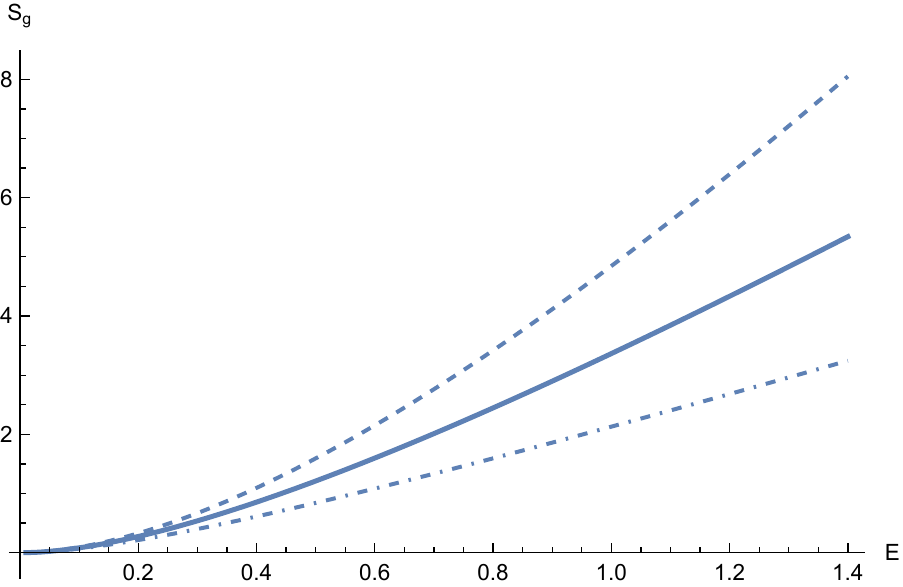}
	\caption{\label{entropy} The entropy $\hbar S_g$ for Schwarzschild spacetime as a function of $E_g$.  The (dashed, solid, and dot-dashed) curves are for $g=1/2$, $g=1$ and $g=2$, respectively. To obtain this plot, the interpolation function in Mathematica was used to obtain $r(E_g)$ and $S_g[r(E_g)]$ was plotted. }
\end{figure}

To define an entropy function with the first law $dE_g = T_g dS_g$ we can compare the differential $dE_g$ to local temperature, the change in entropy (and work, if relevant). The temperature is well-studied.  For instance, Unruh-DeWitt detectors responding to fields in Hartle-Hawking and Unruh (modeling mode behavior expected for a collapsing star) states lead to a thermal response in Schwarzschild spacetime \cite{HLO}.   Static detectors have a thermal response with temperature $T_{loc} = T_H /V$ where $T_H = \hbar (8 \pi M)^{-1}$ \cite{HarHaw,HLO}. Hence,
\begin{equation}
	\label{Stemp}
	T_g = \frac{\hbar \kappa}{2 \pi} (1 - 2M/r)^{-1/2} = \frac{\hbar g}{8 \pi} \left( \frac{r}{M} \right)^2 = \frac{ \hbar \, g}{8 \pi} \left[ gr \left( \sqrt{1+(gr)^2} - gr \right) \right]^{-2}.
\end{equation}
The energy differential may be written as
\begin{equation}
	dE_g = \left( gr dr \right) \left(2 - \frac{gr}{\sqrt{1+(gr)^2}} \right).
\end{equation}
With the temperature in equation (\ref{Stemp}) the entropy differential has the form
\begin{equation}
	dS_g = \frac{dE_g}{T} = \left( \frac{8 \pi r dr}{\hbar} \right) \left[ \frac{(gr)^2}{\sqrt{1+(gr)^2}} \right] \left( 2\sqrt{1+(gr)^2} -gr \right) \left( \sqrt{1+(gr)^2} -gr \right)^2
\end{equation}
which integrates to  
\begin{equation}
	\label{Sentropy}
	S_g = \int_0^r dS_g = \left( \frac{A}{\hbar} \right) (gr)^2  \left[ 1 + 2 (gr)^2 \left( 1 - \sqrt{1+1/(gr)^2} \right) \right],
\end{equation}
where $A= 4 \pi r^2$. As $\epsilon \rightarrow 0$ in $r= 2M(1+\epsilon)$, this entropy goes to $(A_H/4 \hbar)$ where $A_H = 16 \pi M^2$. Thus, the entropy $S_g$ agrees with the Bekenstein-Hawking entropy on the horizon.  (The acceleration $g$ diverges as $1/(4M \sqrt{\epsilon} )$ in this limit.)  The entropy vanishes as $g$ approaches its minimum value of 0. As shown in figure \ref{entropy}, the entropy is a non-decreasing function of the energy $E_g$.  Examples indicate that it satisfies additivity; for two systems with energies $E_{g_1}$ and $E_{g_2}$ satisfying $E_g = E_{g_1}+ E_{g_2}$ then $S_g \geq S_{g_1} +S_{g_2}$.

As in the BTZ case, the expression for the energy differential $dE|_r = dm|_r/V$ integrating from $m=0$ to $m=M$ yields\footnote{As pointed out in \cite{BLY} for more general situations with local matter density $\rho$ and black hole mass $M$ the expression of the energy is the same but now the mass is given by 
\[
	M(r) = 4 \pi \int_0^r \rho(\bar{r}) \bar{r}^2 d \bar{r} + M.
\]
}  \cite{makela}
\begin{equation}
	\label{intE}
	E_{r} = \int_0^M \left( 1 - \frac{2 m}{r} \right)^{-1/2} dm = r \left( 1 - \sqrt{1-\frac{2M}{r}} \right),
\end{equation}
which is the Brown-York quasi-local energy $E_{BY}$ of equation (\ref{EBY}) in Schwarzschild spacetime \cite{BY}. This is the quasi-local energy for non-constant acceleration at fixed areal radius.

In summary, one may construct thermodynamic-like quantities $E_g$ and $S_g$ for the accelerating observers in Schwarzschild spacetimes. The entropy, derived from the first law, has other thermodynamic properties.  There is no independent check - as far as I am aware - in the literature on the validity of the expression for the entropy, equation (\ref{Sentropy}).  Possible work terms are discussed in appendix \ref{pressure}.

\section{Kottler or Schwarzschild-de Sitter spacetimes}
\label{kottler}

Adding a positive cosmological constant $\Lambda$ results in the Kottler \cite{kottler} or Schwarzschild-de Sitter (SdS) spacetimes with metric
\begin{equation}
	ds^2 = - N^2 dt^2 + N^{-2} dr^2 +r^2 d \Omega_2^2
\end{equation}
where
\[
 	N = \left( 1 - \frac{2 M }{r}  - \frac{\Lambda r^2}{3} \right)^{1/2}.
\]
When $0<M< \left( 3 \sqrt{\Lambda} \right)^{-1}$ there are two physical horizons, one black hole horizon
\begin{equation}
	r_H = \frac{2}{\sqrt{\Lambda}} \cos \left[ \frac{1}{3} \arccos\left( - 3 M \sqrt{\Lambda} \right) + \frac{ 4 \pi}{3} \right]
\end{equation}
and one cosmological horizon
\begin{equation}
	r_\Lambda  = \frac{2}{\sqrt{\Lambda}} \cos \left[ \frac{1}{3} \arccos\left( - 3 M \sqrt{\Lambda} \right) \right],
\end{equation}
corresponding to the two physical roots of the lapse.

The magnitude of the acceleration of the CAOs is
\begin{equation}
	g = \frac{M/r^2 - \Lambda r /3}{\sqrt{1 - 2 M/r - \Lambda r^2/3}}. 
\end{equation}
The acceleration vanishes when the gravitational reach of the black hole dissipates, when $r_0=(3 M/\Lambda)^{1/3}$. This radius can be surprisingly ``close" to the black hole.  For instance, for a solar mass black hole and a background cosmological constant of our universe, $\Lambda \simeq 1\times 10^{-52}$ m$^{-2}$, the radius is $r_0 \simeq 350$ lyr, only a couple orders of magnitude beyond the sun's gravitational neighborhood \cite{alex}.   Outside this radius the de Sitter background dominates. This makes sense in that gravitational effects of the black hole fade into the noise on the scale of the next gravitational structure, such as a galaxy or wider cosmological setting. For the region $r>r_0$, the CAOs would be accelerating toward, rather than away from, the black hole.  The infinite acceleration surface is at the cosmological horizon. As the observers approach this surface the local temperature would be that of the radiation from the cosmological horizon. For these reasons, the following analysis is confined to the exterior region with radius, $r_H < r < r_0$. 

Proceeding with the derivation of energy as in the Schwarzschild case, the mass may be expressed as a function of $gr$ and $\Lambda$,
\begin{equation}
	\label{massK}
	M = gr^2\left[ \sqrt{1 +(gr)^2(1-\Lambda/g^2)} - gr \left(1- \frac{\Lambda}{3 g^2} \right) \right].
\end{equation} 	
The energy differential is
\begin{equation}
	dE_g = \left( \frac{ g r}{1 - r^3 \Lambda/ (3M) } \right) \left( \frac{ 2r - 3M - 2 r^3 \Lambda + r^4 \Lambda / (3M) }{r - M - 2 r^3 \Lambda / 3} \right) dr,
\end{equation}
which simplifies to
\begin{equation}
	dE_g = \left( gr dr \right) \left( 2 - \frac{gr (1- \Lambda/g^2) }{\sqrt{ 1 +(gr)^2(1- \Lambda/g^2)}} \right),
\end{equation} 
using the solution for the mass in equation (\ref{massK}). Integration is essentially the same as the Schwarzschild case, but here $gr \sqrt{1 - \Lambda/g^2} = \tilde{u} = \sinh x$ is a better choice for the integration of the second term. The result is
\begin{equation} 
	\label{intEK}
	E_{g} =  g \frac{ A}{4 \pi} - \frac{1}{2g} \left[ \frac{\ln \left(gr \sqrt{1 - \Lambda/g^2} - \beta \right)}{\sqrt{1 - \Lambda/g^2}} + gr \beta \right],
\end{equation}
with 
\[
	\beta = \sqrt{1+(gr)^2(1-\Lambda/g^2)}.
\]
Again, the energy is expressed in terms of quantities accessible to CAOs, acceleration, area, and cosmological constant.

In the large acceleration limit 
\[
	E_g \simeq \frac{g A}{8 \pi} \left(1+ \frac{\Lambda}{2g^2} \right) + O\left(\frac{\ln(g^2A)}{g} \right).
\]
For small cosmological constant, the low acceleration limit of the energy can readily obtained with a double expansion,  
\begin{equation}
	E_g \simeq \frac{g A}{4 \pi} \left[ 1 - \frac{gr}{3}\left( 1 - \frac{\Lambda}{g^2} \right) \right],
\end{equation}
which goes to $M$ as $g \rightarrow 0$ when $r \rightarrow r_0$. 

As in the previous examples the direct integration of $dE_r = dM|_r/V$ with respect to mass and neglecting the constancy of $g$ yields
\begin{equation}
	E_{r} = r \left( \sqrt{1 - \frac{r^2 \Lambda}{3}} - \sqrt{1 - \frac{2M}{r} - \frac{r^2 \Lambda}{3}}  \right),
\end{equation}
which is the Brown-York quasi-local energy with respect to a de Sitter background. 

Due to the de Sitter background the derivation of the entropy is more subtle in this case, in part due to the lack of well-defined physical temperature and well-developed effective asymptotics for Kottler spacetimes.  Even if the temperature is again of the form $T_{loc} = T_H/V$ then there remains an additional question about the norm of the Killing vector $\xi^a$. There are two obvious possibilities: One may consider the Killing vector field $\xi^\mu = (\partial_t)^\mu$, as one would naturally use in the asymptotically flat case \cite{GH}. This yields a surface gravity of 
\begin{equation}
	\kappa = \frac{1}{2} N'|_{r_H} = \frac{1}{2} \left( \frac{1}{r_H} - r_H \Lambda \right).
\end{equation}	
Alternatively, one may note \cite{BH} that there is an ``asymptotically flat-like" region in that at $r_0$ the acceleration vanishes and the Killing vector field 
\begin{equation}
	\tilde{\xi}^\mu = \left. \frac{1}{N} \right|_{r_0} \left( \partial_t \right)^\mu
\end{equation}
is geodesic in this region. Let's consider the first case and assume that in the relevant region, $r_H < r < r_0$ the temperature takes the redshifted Hawking form, $T_{g} = T_H/V$.
Due to presence of $r_H$ in the temperature, which is a function of both $g$ and $r$ via the mass, the integration of entropy defined by $S = \int dE/T$ is not as straightforward as the previous cases. The second choice of the normalization of the Killing vector field results in an additional factor that is independent of $g$ and $r$.

Finally a quick comment on the anti-de Sitter case, when  $ \Lambda = -3/\ell^2 <0 $ and there is only one black hole horizon, 
\begin{equation}
	r_H = \left( M \ell^2 \right)^{1/3} \left[ \left( 1+ \sqrt{1+\frac{\ell^2}{27 M^2}} \right)^{1/3} + \left( 1- \sqrt{1+\frac{\ell^2}{27 M^2}} \right)^{1/3}\right].
\end{equation}
The acceleration does not vanish at finite radius.  However, the calculation of the energy proceeds as before, with the same result (\ref{intEK}).  The temperature is known to be of redshifted Hawking form \cite{HP,NHLMM}.  As in the de Sitter case the horizon radius $r_H$ in the surface gravity makes the calculation of the entropy less straightforward.

\section{Discussion}

In their work on energy for observers perched in the near-horizon geometry of black holes, Frodden, Ghosh and Perez \cite{FGP} and M\"akel\"a \cite{makela} found remarkably simple expressions; energy was proportional to area. This initial exploration was motivated by asking whether this expression generalizes to the exterior region. It seems to, although the expressions are not quite as simple.

Often we say that a definition of energy exists when a mass-dimension quantity is conserved under dynamical evolution.  Here the perspective is different: The energy is measured by CAOs, which are not undergoing geodesic motion, as the mass of the black hole changes. The energy tracks the flow of stress-energy observed by these observers. The definition of energy $E_g$ for constantly accelerating observers is extended to the exterior regions in BTZ, Schwarzschild, and Kottler spacetimes.  In every case these energies are expressed in terms of geometric quantities accessible to these observers. The expressions of the energy appear to be novel in that, instead of integrals of 2-forms on a surface, these are integrals over the flow of energy as the central mass is formed.  As the energy $E_g$ is constructed from local geometric quantities accessible to observers, it might be useful in constructing quasi-local descriptions of these spacetimes. Additionally in the BTZ and Schwarzschild cases, candidate expressions for the entropies associated to these observers are defined.

All these expressions depend on the existence of a normed timelike Killing vector field.  For the energies to be meaningful, the Killing field must have unit norm, and thus be the 4-velocity of geodesic observers, in some (asymptotic) region of the spacetime.   While clearly the case for black hole spacetimes with flat asymptotic geometries, this has to be more carefully defined for the asymptotically de Sitter and anti-de Sitter spacetimes. 

One way of testing these quantities is in (quasi-)local expressions of laws analogous to black hole thermodynamics. By analogy we would expect that these definitions should result in the following:
\begin{enumerate}
	\item The temperature $T_g$ is constant on the surface containing the observers. In the cases considered here this is trivial by choice of CAOs.
	\item First law: $dE_g = T_g dS_g + \text{ work terms}$.  This was used to find candidate expressions for the entropy $S_g$.
	\item Second law: $d S_g \geq 0$, which is as yet largely unexplored.
	\item Third law: As the acceleration decreases to its minimum value determined by ambient curvature or 0, the entropy settles to $S=0$ (or to the minimum value of the frozen-in residual entropy) as shown in the non-rotating BTZ and Schwarzschild cases.
\end{enumerate}
Although there are indications that in the case of the non-rotating BTZ and Schwarzschild spacetimes that these may hold, further study is required.

\bigskip

\noindent{\large\bf Acknowledgements} it is a pleasure to thank Jorma Luko, Jormo M\"akel\"a, and Andrew Projanski for enlightening conversations. 

\appendix

\section{Normalizing $E_{loc}$ on the horizon}
\label{Enorm}
In the context of general relativity, it is well known that if there is a timelike Killing vector field $\xi_a$ then there is a conserved energy flux $\epsilon^a = - T^a_b \xi^b$. On a Cauchy surface $\Sigma$ with normal $n_a$, the associated energy 
\[
	{\mathcal E}  =  \int_\Sigma \epsilon^a n_a d^3x
\]
is independent on the choice of $\Sigma$. This has many of the desired properties we would like e.g. for an isolated gravitational body, the energy  ${\mathcal E}$ agrees with the asymptotic mass $M$, here I focus on CAOs. For a fleet of (constantly accelerating) observers ${\mathcal O}$ on a black hole spacetime with timelike Killing vector field $\xi^a$, 
radially propagating test matter-energy with stress-energy tensor $\delta T^{ab}$ has the conserved energy flux $\varepsilon^a = - \delta T^a_b \xi^b$.  The worldsheet of the observers ${\mathcal W}_{\mathcal O}$ divides the external part of the spacetime into two regions, ${\mathcal R}_1$ and ${\mathcal R}_2$, where ${\mathcal R}_1$ is between the horizon ${\mathcal H}$ and the worldsheet of the observers ${\mathcal W_{\mathcal O}}$ and ${\mathcal R}_2$ is bounded by the worldsheet and spatial infinity. The spatial hypersurfaces bounding the regions to the past and future are $\Sigma_1$ and $\Sigma_2$, respectively. 

Given an energy flux with compact support it is always possible to find a spacetime region ${\mathcal R}_1$ such that $\varepsilon^a |_{\Sigma_i} =0$ so 
\[
	\int_{\mathcal W_{O}}  \varepsilon^a d \Sigma_a =  \int_{\mathcal H}  \varepsilon^a d \Sigma_a.
\]
Or,
\begin{equation}
\label{firstlaw}
	\int_{\mathcal W_{O}}  \delta T^{ab} \xi_a n_b d \tau dS =  \int_{\mathcal H}  \delta T^{ab} \xi_b k_a dv dS,
\end{equation}
where $\tau$ is the observers' proper time, $k_a$ is the null normal to the horizon satisfying $k^b \nabla_b k^a = \kappa_{\mathcal H} k^b$ with parameter $v$, $n_a$ is the inwardly pointing normal on the observers' spatial 2-surface ${\mathcal S}$, and $dS$ is the measure on the spatial surfaces. 

For the observers ${\mathcal O}$ the measured change in energy is
\[
	\delta E_{loc} =  \int_{\mathcal W_{O}}  \delta T^{ab}  u_a n_b d \tau dS.
\] 

The observers' 4-velocity is related to the Killing vector field via the redshift factor $V$, 
$ u^a = \xi^a/V$. For example in Schwarzschild spacetime,
\[
	| \xi | = \sqrt{- \xi^\alpha \xi_\alpha} = V \text{  with } V = \sqrt{  1 - \frac{2 M}{r}}.
\]
While this norm is not constant all along the worldsheet $\mathcal{W_O}$, I  work at first order in the change due to the addition of the test stress-energy $\delta T^{ab}$. 
At leading order  the energy may be written as
\[
	\int_{\mathcal W_{O}}  \delta T^{ab}  \xi_a n_b d \tau dS \simeq V \int_{\mathcal W_{O}}  \delta T^{ab}  u_a n_b d \tau dS = V \delta E_{loc}.
\]
Comparing stationary states before and after the perturbation and using Raychaudhuri's equation at first order the usual arguments (e.g. see Ref. \cite{poisson}) give
\[
	\int_{\mathcal H}  \delta T^{ab} \xi_b k_a dv dS \simeq \frac{\kappa_{\mathcal H}}{8 \pi} \delta A_{\mathcal H} 
\]
So,
\begin{equation} 
	\delta E_{loc} \simeq \frac{1}{V} \frac{\kappa_{\mathcal H}}{8 \pi} \delta A_{\mathcal H} = \frac{1}{V}  \delta M,
\end{equation}
where the surface gravity and area expressions were used. This expression relates the change in the observers' quasi-local energy and the parameter $M$ that characterizes the spacetime. When the spacetime has an asymptotically flat region, this argument works in a similar manner for the space time region ${\mathcal R}_2$.  The form of $E_{loc}$ is the same as equation (\ref{Eloc}). 

\section{GEMS circumference calculation for accelerating observers} 
\label{EGEMS}
Following \cite{carlip95rev}, three dimensional anti de-Sitter spacetime, AdS$_3$,  can be expressed as a submanifold in the higher dimensional space $\mathbb{R}^{2,2}$ with coordinates $(T_1,T_2,X_1,X_2)$ and two-time metric
\[
	dS^2 = - dT_1^2 - dT_2^2 + dX_1^2 +dX_2^2 
\]
when 
\[
	-T_1^2 - T_2^2 + X_1^2 + X_2^2  = -\ell^2. 
\]
The BTZ exterior geometry of equation (\ref{btzmetric}) arises as a quotient of an open region in adS$_3$ by the discrete group $\mathbb{Z}$ realized in the identification $(t,r,\varphi) \sim (t,r, \varphi+2 \pi)$ when,
\begin{equation}
	\begin{split}
	T_1 &= \ell \sqrt{\alpha} \cosh \left( \frac{r_+}{\ell} \varphi - \frac{r_-}{\ell^2} t \right),\\
 	T_2 &= \ell \sqrt{\alpha-1} \sinh \left( \frac{r_+}{\ell^2} t - \frac{r_-}{\ell} \varphi \right),\\
	X_1 &= \ell \sqrt{\alpha} \sinh \left( \frac{r_+}{\ell} \varphi - \frac{r_-}{\ell^2} t \right),\\
	X_2 &= \ell \sqrt{\alpha-1} \cosh \left( \frac{r_+}{\ell^2} t - \frac{r_-}{\ell} \varphi \right),
	\end{split}
\end{equation}
where
\begin{equation}
	\alpha = \frac{r^2 - r_-^2}{r_+^2-r_-^2} =  \left(1 - 1/(g\ell)^2 \right)^{-1}.
\end{equation}
As computed in \cite{DL},  the entropy is the area of the ``transverse" surface 
\begin{equation}
	T_1^2 - X_1^2 = \ell^2 \alpha
\end{equation}
with the appropriate periodicity condition.  The calculation done here includes rotation, when the observer is co-rotating with the horizon, but the result is the same when there is no rotation. Choosing a strip between $-\pi$ and $\pi$ one can compute the area $I$ on this surface at constant acceleration,
\begin{equation}
	I = \int dT_1 dX_1 \, \delta\left( \sqrt{T_1^2-X_1^2} - \ell \sqrt{\alpha} \right).
\end{equation}
Using the $\delta$-function to eat the $T_1$ integration one finds
\begin{equation}
	\label{btz_circum}
	I = \int_{\ell \sqrt{\alpha} \sinh \left(-r_+ \pi/\ell - r_-t/\ell^2 \right)}^{\ell \sqrt{\alpha} \sinh \left(r_+ \pi/\ell - r_-t/\ell^2 \right)} dX_1 \frac{\ell \sqrt{\alpha}}{\sqrt{\ell^2 \alpha +X_1^2}} = 2 \pi r  \sqrt{\alpha}.
\end{equation}
This ``circumference" $I$ gives the result in equation (\ref{btz_area}). The calculation here generalizes the one in \cite{DL} to any exterior radius.

\section{Pressure terms}
\label{pressure}

Possible work terms in the first law may be explored with the expected boundary terms in the Brown-York construction \cite{BY,bcm}.  In the BTZ case the surface pressure is given by \cite{bcm}
\begin{equation}
 	P = \sigma_{ab}s^{ab} = \sigma_{\varphi \varphi}s^{\varphi \varphi} = \frac{g}{\pi} + \frac{\partial \left( r \epsilon_o \right) }{\partial r} 
\end{equation}
where $\sigma_{ab}$ is the metric on $\partial \Sigma$, $s^{ab}$ is the surface pressure density, and $\epsilon_o$ is the reference energy surface density.  Were the black hole to decay, the spacetime would be adS$_3$ so it seems clear that anti-de Sitter is the correct background, $\epsilon_o = - \sqrt{1 +r^2/\ell^2} / (\pi r)$ \cite{bcm}.  Choosing reference observers accelerating at the same magnitude $g$ in adS$_3$, one finds that the reference term is $- g/\pi$ and so the surface pressure vanishes in this case.

For Schwarzschild black holes there is a non-vanishing work term $-PdA$ derived in \cite{BY} yielding a pressure of 
\begin{equation}
	P = \left( \frac{1}{8 \pi r} \right) \left( \frac{1 - M/r}{\sqrt{1 - 2M/r}} -1 \right),
\end{equation}
when the reference spacetime is Minkowskian. The work term becomes
\begin{equation}
	- P dA = \left( g r dr \right) \left( 1 - r/M  + 1/(gr) \right)
\end{equation}
for the accelerating observers. Including this in a first law, however, results in an entropy with incorrect limits both at high and vanishing accelerations.  Furthermore, unlike the case in \cite{BY} where the work term scales with $dA$ and the entropy scales as $dM$, the work and entropy terms are redundant in that they both scale as $dA$. 

\section{Expressions for $E_g$}
\label{altEgS}
The energy in the Schwarzschild case can be expressed in a variety of ways,
\begin{equation} \label{intE3}
	\begin{split}	
	E_{g} &= \int_0^{r_g(M)} \sinh x \left(1 + \frac{e^{-x}}{\cosh x} \right) dr \\ 
	&=  g \int_0^{r_g(M)} r dr + 
	\frac{1}{g} \int_o^{x_g(M)} e^{-x} \sinh x \, dx \\
		&= g \frac{ A}{8 \pi} - \frac{1}{2g} \ln V - \frac{1}{2g} \frac{M}{r} \\
		&= g \frac{ A}{8 \pi} - \frac{1}{2g} \ln V  - \frac{V \sqrt{A}}{4 \sqrt{\pi} }\\
		&= g \frac{ A}{8 \pi} + \frac{1}{4} \sum_{n=2}^\infty \left( \frac{g^{n-1}}{n} \right) \left[2 r \left( \sqrt{1+(g r)^2} - g r \right) \right]^n.
	\end{split}  
\end{equation} 
These are equivalent to the expression in equation (\ref{intE2}). The splitting of the integral this way organizes the divergence as the observers approach the horizon.  Letting $r = 2M(1+\epsilon)$ the terms scale with $\epsilon$ as $\epsilon^{-1/2}$, $\epsilon^{1/2} \ln \epsilon$, and $\epsilon^{1/2}$, respectively.  The last term, $-M/2gr$, can be also expressed as (1/2) the `Tr$K$' portion of the Brown-York quasi-local energy, $-r \sqrt{1- 2M/r}$.

\end{document}